# Topological transition and emergent elasticity of dislocation in skyrmion lattice: Beyond Kittel's magnetic–polar analogy

*Kohta Kasai[1]\*, Akihiro Uematsu[1], Tatsuki Kawakane[1], Yu Wang[2], Tao Xu[1], Chang Liu[1], Susumu Minami[1], Takahiro Shimada[1]\**

[1]Department of Mechanical Engineering and Science, Kyoto University; Nishikyo-ku, Kyoto 615-8540, Japan.

[2]School of Civil Engineering, Central South University, Changsha, 410083, China.

\*Corresponding author.
Email: kasai.kohta.72s@st.kyoto-u.ac.jp, shimada.takahiro.8u@kyoto-u.ac.jp

# Abstract

Magnetic and polar skyrmions exhibit topologically protected quasiparticle behavior, including emergent fields, deformation, and the formation of a densely packed skyrmion lattice, beyond conventional domain configurations described by Kittel's law. Analogous to atomic crystals, lattice defects, especially dislocations and their associated strain fields, are crucial for understanding the lattice behavior of skyrmions; however, their features and roles remain insufficiently understood. Here, we show that magnetic skyrmion dislocations develop a core-split structure due to a significant skyrmion elongation up to 180% of their original length, reaching a topological transition from a single skyrmion to two half-skyrmions. Despite such a distinct structure, the long-range strain fields around the dislocation perfectly obey conventional Volterra's elasticity theory, in contrast to polar skyrmion lattices, where skyrmion deformations cause a breakdown of the elasticity theory. Furthermore, an energetic analysis shows that Dzyaloshinskii–Moriya interaction drives the large skyrmion deformation of the dislocation core. Our findings not only clarify the coexistence of topological core-reconstruction and a robust long-range elastic field of dislocations in magnetic skyrmion lattices, but also reveal that magnetic and electric domains, long regarded as dual and analogous, exhibit fundamental differences when extended into the regime of collective topological quasiparticles.

# Introduction

Magnetic skyrmions are deformable quasiparticles stabilized by topological protection, and have attracted considerable attention owing to their unusual physical properties[1–4]. For example, the emergent electromagnetic fields of skyrmions allow them to be driven at high speed with small currents via the topological Hall effect and skyrmion Hall effect[5–7]. Such unique physical properties have led to proposals to apply skyrmions to nanodevices such as racetrack memories[8], logic elements[3,9], and synapses in neuromorphic computing[4,10]. Furthermore, beyond their importance in magnetism, the concept of skyrmions has been extended to a wide range of physical systems, including acoustic[11,12], optical[13], and polar skyrmions[14,15]. Among them, magnetic skyrmions remain the primary focus, while polar skyrmions have been reported as their electric counterparts, highlighting the magnetic–electric duality. Unlike conventional magnetic and polarization domains, which are described by Kittel's law[16,17] as continuous domain structures, skyrmions exhibit discrete particle-like behavior as quasiparticles[18–21].

The most striking manifestation of this discrete nature is the formation of skyrmion lattices[1,2,21–24], in which individual skyrmions act as the constituent particles of a crystalline array. Similar to atomic crystals or other two-dimensional lattices, skyrmion lattices exhibit collective behaviors such as lattice rotation[25,26], melting[27], and other phenomena[23,28], clearly demonstrating their crystalline nature. Furthermore, like their atomic counterparts, skyrmion lattices contain lattice defects such as dislocations[29–32], vacancies[20,33], and grain boundaries[34–37]. In general, such lattice defects play a crucial role in determining a material's macroscopic mechanical properties. In particular, dislocations and their motion drive key mechanical processes such as hardening, fatigue, and plastic deformation[38,39]. In skyrmion lattices, crystal rotations can be mediated by dislocations composing grain boundaries[25]. Moreover, in the context of a two-dimensional phase transition based on KTHNY (Kosterlitz, Thouless, Halperin, Nelson, and Young) theory, the emergence of numerous dislocations plays a central role in the melting and freezing of skyrmion lattices[27,31,40–42]. These studies indicate that, even in skyrmion lattices, which are emergent quasiparticle lattices distinct from conventional crystals, dislocations are also essential for understanding their overall behavior as skyrmion lattices. On the other hand, as a feature specific to skyrmion lattices, several studies reported skyrmion deformations near lattice defects. For example, it has been theoretically predicted that a skyrmion vacancy exhibits large deformations of surrounding skyrmions, leading to structural changes depending on external conditions[33], and experimental observations have also reported shape distortions of skyrmions located near dislocations[30,34,43]. However, the quasiparticle-specific structural properties of magnetic skyrmion dislocations have not been investigated sufficiently.

Another critical aspect of the dislocations is the induced strain field. In atomic crystals, dislocations generate continuous elastic fields around dislocation cores described by Volterra's elasticity theory[44]. In magnetic skyrmion lattices, although an elastic-like strain field around a dislocation core has also been reported[29], this observation is limited to conditions where skyrmion deformation is negligibly small, and each skyrmion can be approximated as a rigid particle. On the other hand, in the case of polar skyrmions, large skyrmion deformation fundamentally alters lattice mechanics, leading to the breakdown of conventional elasticity and associated strain fields[45]. A similar transition in lattice mechanics may also occur in magnetic skyrmions when the deformation becomes significant; however, this possibility remains unexplored.

Here, we demonstrate detailed magnetic skyrmion dislocation structures and their inherent strain fields. Our phase-field simulations reveal the quasiparticle-specific dislocation core structures, accompanied by significant skyrmion deformation, and their dependence on magnetic field and temperature. Quantitative strain analysis establishes the mechanical characteristics of magnetic skyrmion dislocation. Furthermore, comparison with polar skyrmion dislocations reveals fundamental differences in the mechanical properties of the skyrmion lattice between magnetic and polar systems, long regarded as dual and analogous.

## Results and Discussion

We investigated skyrmion dislocation structures in magnetic skyrmion lattices in MnSi thin films using phase-field simulations (see Methods). Initially, the skyrmions were arranged in a triangular lattice with ideal dislocations. The skyrmion spacing $a$, which depends on the magnetic field and temperature, was obtained from the literature[33]. The time evolution of magnetic moment distribution was calculated until an equilibrium state was reached to obtain the stable dislocation structure. Figure 1 shows the magnetic moment distribution near the dislocation core under the conditions of $T = 22.5$ K and $H = 2.0 \times 10^5$ A/m. As shown in Figure 1(a), the skyrmion lattice stabilizes while maintaining a single isolated dislocation. Each skyrmion has a Bloch-type configuration with in-plane and out-of-plane components of the magnetic moment vector, consistent with typical skyrmions in ferromagnetic materials. In the vicinity of the dislocation core (Figure 1(b)), the core comprises a 5-fold skyrmion, neighboring five skyrmions, and a 7-fold skyrmion, neighboring seven skyrmions, forming a 5-7 pair structure commonly observed in two-dimensional triangular lattices. Morphologically, the 5-fold skyrmion contracts while the 7-fold skyrmion elongates slightly, whereas the other skyrmions remain nearly circular. These deformations correspond to the compressive and tensile strain fields of typical dislocations and are consistent with the experimental observations of skyrmion dislocations[34,43], supporting the validity of our simulations.

The degree of skyrmion deformation depends on external fields, such as the magnetic field and temperature[33,46]. To investigate how skyrmion deformations affect dislocation structures, we performed simulations at different magnetic field strengths at a fixed temperature ($T = 22.5$ K) within the stable skyrmion phase. Under a high magnetic field ($H = 3.5 \times 10^5$ A/m), the same 5-7 pair structure observed at $H = 2.0 \times 10^5$ A/m is obtained, while each skyrmion becomes smaller. The variation in shape among skyrmions is also smaller than that under $H = 2.0 \times 10^5$ A/m (Figure 2(a), (b)). In contrast, under a low magnetic field ($H = 1.0 \times 10^5$ A/m), skyrmions near the dislocation core undergo pronounced deformations, particularly in the 5-fold and 7-fold skyrmions (Figure 2(c)). The length of the 7-fold skyrmion increases to 180% of the original value along the $x_1$ direction, and adjacent skyrmions are displaced away from the dislocation core. Meanwhile, the 5-fold skyrmion shrinks to 90% of its original size. Despite such large deformations, the relative arrangement of neighboring skyrmions is maintained, and the 5–7 dislocation pair is preserved, resulting in a dislocation structure unique to skyrmions with quasiparticle nature. Thus, skyrmion dislocations exhibit field-dependent structures: conventional atom-like structures with slight skyrmion deformations under high magnetic fields, and quasiparticle-specific structures with large deformations under low magnetic fields.

To capture the topological features of these skyrmions, we calculated a Pontryagin charge (topological charge) density $q = \boldsymbol{m}(\partial \boldsymbol{m}/\partial x_1 \times \partial \boldsymbol{m}/\partial x_2)$, where $\boldsymbol{m}$ is the normalized magnetic moment vector[47]. The integral of Pontryagin charge density is an invariant known as the skyrmion number. As shown in Figure 2(a-2)(b-2), negative charge density values are distributed in each skyrmion, indicating the existence of an isolated single skyrmion. However, the elongated 7-fold (core-site) skyrmion at a low field exhibits a different distribution (Figure 2(c-2)). Negative values appear on both sides, while positive values emerge at the center, suggesting the onset of a topological transition. Figure 2(d) shows the detailed magnetic moment configuration of the elongated 7-fold skyrmion. The structure consists of two half-skyrmions (merons)[48,49] on opposite sides, connected by a stripe phase region, consistent with the topological charge distribution. Thus, the 7-fold skyrmion has undergone a transition from a single skyrmion into two split half-skyrmions (merons) linked by a stripe phase. In a typical skyrmion lattice, the lattice is defined by a bijective projection between lattice points and individual skyrmions[35,50], meaning that each skyrmion center corresponds to one lattice point. However, in the present case of the combination structure, each half-skyrmion possesses its own skyrmion center, effectively introducing an additional lattice point. As a result, around the dislocation core, the number of neighboring lattice points is altered, and the dislocation core is reconstructed (Figure 2(e). The lattice point associated with one of the half-skyrmions becomes a newly formed 5-fold point, while the lattice point located below it becomes a 7-fold point. Consequently, the dislocation core shifts downward by one lattice unit from its original position. Therefore, from a topological viewpoint, the significant skyrmion elongation induces a core-split dislocation structure and causes a positional change in the dislocation core. These results highlight that skyrmion lattices, as assemblies of deformable quasiparticles, can accommodate the lattice defects in a way not possible in conventional atomic crystals.

Temperature also modulates the skyrmion deformability. Additional calculations at varied temperatures with a fixed magnetic field ($H = 2.0 \times 10^5$ A/m) show that the size and deformations are small at high temperatures and large at low temperatures, though less sensitive than the magnetic field effect (see Supplemental Material). An $H$-$T$ map summarizing these dependencies is presented in Figure 3. Across all conditions, the 7-fold skyrmion exhibits the most substantial deformation, with maximum elongation at $H = 1.0 \times 10^5$ A/m and $T = 22.5$ K. Overall, skyrmion size and deformation decrease with increasing magnetic field and temperature, and increase under lower field and temperature. These field- and temperature-dependent structural characteristics are also observed in polar skyrmion dislocations, highlighting the close parallels between the magnetic and polar skyrmions.

To elucidate the long-range lattice characteristics of skyrmion dislocations, we focus on the strain fields inherently associated with dislocations. These strain fields are critical because they determine the mechanical and dynamic roles of dislocations, which in turn govern plastic deformation in materials. In classical elasticity, dislocations generate strain fields according to Volterra's elasticity theory[44], which follows a $1/r$ law, where $r$ is the distance from the dislocation core. However, because skyrmions are deformable quasiparticles, conventional elasticity theory may not apply. In polar skyrmion dislocations, for example, strain fields deviate markedly from classical elastic fields depending on the degree of skyrmion deformations[45]. To visualize and evaluate the strain fields of magnetic skyrmion dislocations, we calculated lattice strain, defined as the local deformation measured with respect to a perfect hexagonal skyrmion lattice[45,51]. Figure 4(a)(b) shows the lattice strain fields near the dislocations under high and low magnetic field conditions. As shown in Figure 4(a), under a high magnetic field, where skyrmion deformations are small, continuous elastic-like strain fields appear, closely resembling those in atomic crystals. These results are consistent with experimental observation[29]. It is worth noting that skyrmions act as rigid particles and exhibit elastic collective behaviors like atomic crystals, even though they are quasiparticles composed of magnetic moment vector fields. Furthermore, even under a low field, as shown in Figure 4(b), similar strain fields emerge, except in the immediate vicinity of the dislocation core, despite significant skyrmion deformation. To precisely evaluate these features, Figure 4(c) compares the strain values for various dislocations and the theory. The results show that the strain distributions in magnetic skyrmion dislocations are perfectly consistent with those of atomic dislocations and Volterra's theory, in contrast to the deviations in polar skyrmion dislocation. Therefore, the strain fields associated with magnetic skyrmion lattice dislocations robustly follow conventional elasticity theory, regardless of the external condition-dependent large skyrmion deformation and quasiparticle-specific dislocation structure. Previous studies have primarily investigated the elastic behavior of skyrmion lattices as a response to external physical fields such as mechanical stress and magnetic fields, and have discussed the applicability of Hooke's law[28,52,53]. In this context, the present results reveal the emergence of Volterra's elasticity in skyrmion lattices, thereby extending the conceptual scope of the elasticity theory in these topological quasiparticle lattices.

To clarify the significance of the emergence of Volterra's elasticity in magnetic skyrmion lattices, we compare their behavior with that of polar skyrmion lattices. The robust elasticity in magnetic skyrmion lattices contrasts sharply with polar skyrmion lattices, where the elasticity theory breaks down owing to large skyrmion deformation. In polar skyrmion lattices, when skyrmions act as soft and deformable particles, part of the lattice frustration can be relaxed by skyrmion deformations rather than positional adjustment of each skyrmion, leading to local concentrations of lattice strain[45,54]. In contrast, magnetic skyrmions are not considered to exhibit such relaxation except in the immediate vicinity of the dislocation core, thereby preserving Volterra's law. This indicates that, even though both magnetic and polar skyrmions are deformable quasiparticles, differences in their individual behaviors result in distinct elastic properties at the lattice level. Magnetic skyrmions and polar skyrmions have been regarded as dual counterparts, reflecting the correspondence between magnetic and electric domains that both follow Kittel's law in conventional materials[16,17]. Following the discovery of magnetic skyrmions, their polar analogues were also identified, and both share topological protection and similar vortex-like textures despite distinct formation mechanisms and helicities. However, when extended to the regime of dislocation-mediated lattice deformation, a fundamental distinction emerges. While magnetic skyrmion lattices retain elastic behavior consistent with Volterra's theory, polar skyrmion lattices exhibit non-elastic strain fields. In other words, although both systems share topological and structural analogies, their mechanical behaviors as skyrmion lattices are fundamentally different. This reveals that extending magnetic and electric domain concepts to topological quasiparticle lattices leads to a clear divergence in their collective mechanics.

Finally, to clarify the stabilization mechanism under a low magnetic field, where local skyrmion deformation occurs in the vicinity of the dislocation core, Figure 5(a) shows the temporal evolution of the total energy and each free-energy term during the skyrmion deformation. During this process, the exchange energy increases and the Dzyaloshinskii-Moriya interaction (DMI) energy decreases, leading to a reduction in the total energy. The changes in the other free-energy terms are relatively small and can be neglected. This result indicates that the exchange and DMI energies are the dominant factors governing large skyrmion deformation. These two terms act competitively on magnetic moment distributions: exchange energy favors their alignment, corresponding to non-skyrmion regions, while DMI energy favors twisting magnetic moments, corresponding to skyrmion regions. Figure 5(b) supports this interpretation, showing that the spatial distribution of exchange and DMI energies are altered with skyrmion deformation. Before deformation, the periphery of the 7-fold skyrmion corresponds to a non-skyrmion region, exhibiting low exchange energy and high DMI energy. When the skyrmion elongates to occupy this region, the exchange energy increases while the DMI energy decreases. To quantify this behavior, the energy change per dislocation core is calculated as $\Delta F_i = \left( F_{i,\text{after}} - F_{i,\text{before}} \right)/$ $N_{\text{core}}$, where $F_{i,\text{after}}$ and $F_{i,\text{before}}$ are free energies of the system after and before skyrmion deformation, and $N_{\text{core}}$ is the number of dislocations in the model. In this process, $\Delta F_{\text{exchange}} = +6.29 \times 10^{-20}$ J and $\Delta F_{\text{DMI}} = -8.26 \times 10^{-20}$ J are obtained. Since the DMI energy reduction exceeds the exchange energy increment, the skyrmion deformation lowers the total free energy. Therefore, the balance between exchange and DMI energies governs the stability of non-skyrmion regions associated with lattice mismatch around dislocations, ultimately leading to the large deformation of 7-fold skyrmions.

In summary, we have investigated the magnetic skyrmion dislocation structures and their dependence on magnetic field and temperature. Dislocations form typical 5-7 pair structures, while skyrmions around the dislocation core exhibit significant deformation: 5-fold skyrmions tend to shrink, whereas 7-fold skyrmions elongate markedly. Skyrmion deformation increases at lower magnetic fields and lower temperatures, with the elongated skyrmion reaching almost twice the original size. The elongated skyrmion undergoes a topological transition, splitting into two half-skyrmions and inducing a local lattice reconstruction of the dislocation core. As a result, the dislocation core shifts downward by one lattice unit from a topological viewpoint. Despite such a distinct core structure, the dislocation-induced strain field obeys conventional Volterra's elasticity theory, in sharp contrast to polar skyrmion dislocations, where elasticity breaks down owing to skyrmion deformations. Energetic analysis further revealed that the local skyrmion deformation is driven by the Dzyaloshinskii-Moriya interaction energy, which competes with the exchange energy interaction. This study not only clarifies the coexistence of intrinsic robustness of elasticity in magnetic skyrmion lattices but also uncovers a new aspect of magnetic and electric domain organization. Although these two systems have long been regarded as dual and analogous, as exemplified by Kittel's law, they exhibit fundamental differences in the regime of skyrmion lattices.

## Methods

The time evolution of the magnetization $\boldsymbol{M}$ is described by the Time-dependent Ginzburg-Landau (TDGL) equation,

$$\partial M_i / \partial t = -L(\delta F / \delta M_i), \tag{1}$$

where $t$ denotes time, $L$ is the kinetic coefficient, and $F$ represents the total free energy of the system. The mechanical equilibrium equation is imposed through the stress-strain relationship,

$$\sigma_{ij,j} = \left(\partial f / \partial \varepsilon_{ij}\right)_{,j} = 0, \tag{2}$$

where $\sigma_{ij,j}$ denotes the stress gradient and $\varepsilon_{ij}$ is the strain tensor. In addition, the magnetic equilibrium is ensured by Maxwell's equations,

$$B_{i,i} = (-\partial f / \partial H_i)_{,i} = 0, \tag{3}$$

where $B_{i,i}$ and $H_i$ denote the magnetic flux density and magnetic field in the material. Eqs. (1), (2), and (3) govern the evolution of the system in the phase-field simulation. The specific forms of the energy function and computational procedures are provided in the Supplemental Material.

ASSOCIATED CONTENT

**Supporting Information**

Additional materials on phase-field simulation, models, and additional calculations (DOC).

AUTHOR INFORMATION

**Corresponding Authors**

Kohta Kasai

Department of Mechanical Engineering and Science, Kyoto University; Nishikyo-ku, Kyoto 615-8540, Japan.

Email: kasai.kohta.72s@st.kyoto-u.ac.jp

Takahiro Shimada

Department of Mechanical Engineering and Science, Kyoto University; Nishikyo-ku, Kyoto 615-8540, Japan.

Email: shimada.takahiro.8u@kyoto-u.ac.jp

**Author Contributions**

K. K. and A. U performed the phase-field simulations and analyzed the data. K. K wrote the manuscript. T. K., T. X, C. L., Y. W., and S. M. provided critical feedback on the manuscript. T. S. conceived the project and supervised the work. All authors contributed to the general discussion and comment on the manuscript.

ACKNOWLEDGMENT

This work was supported by JSPS KAKENHI (Grant Numbers JP23H00159, JP23K17720, JP24H00032), JST FOREST Program (Grant Number JPMJFR222H), JSPS International Research Fellow (No. P22065), JSPS SPRING (Grant Number JPMJSP2110), National Natural Science Foundation of China (12402194), and the Hunan Provincial Natural Science Foundation of China (2025JJ60053)

# Data availability

All data generated or analyzed during this study are included in this published article, and the datasets used and analyzed during the current study are available from the corresponding author upon reasonable request.

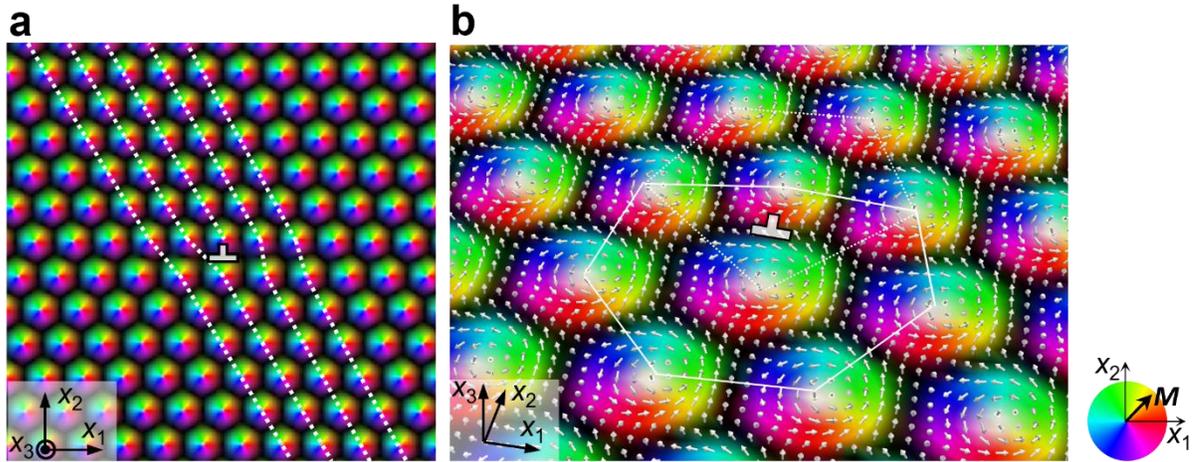

Figure 1. (a) Overview of skyrmion lattice including a skyrmion dislocation. The black area represents the out-of-plane magnetic moment, and the colored areas represent the in-plane magnetic moment, forming skyrmions. The white dashed lines represent the alignment of skyrmions. (b) 5-7 pair structure of the skyrmion dislocation core. The arrows represent magnetic moments, and the white lines represent pentagon and heptagon lattices comprising the dislocation.

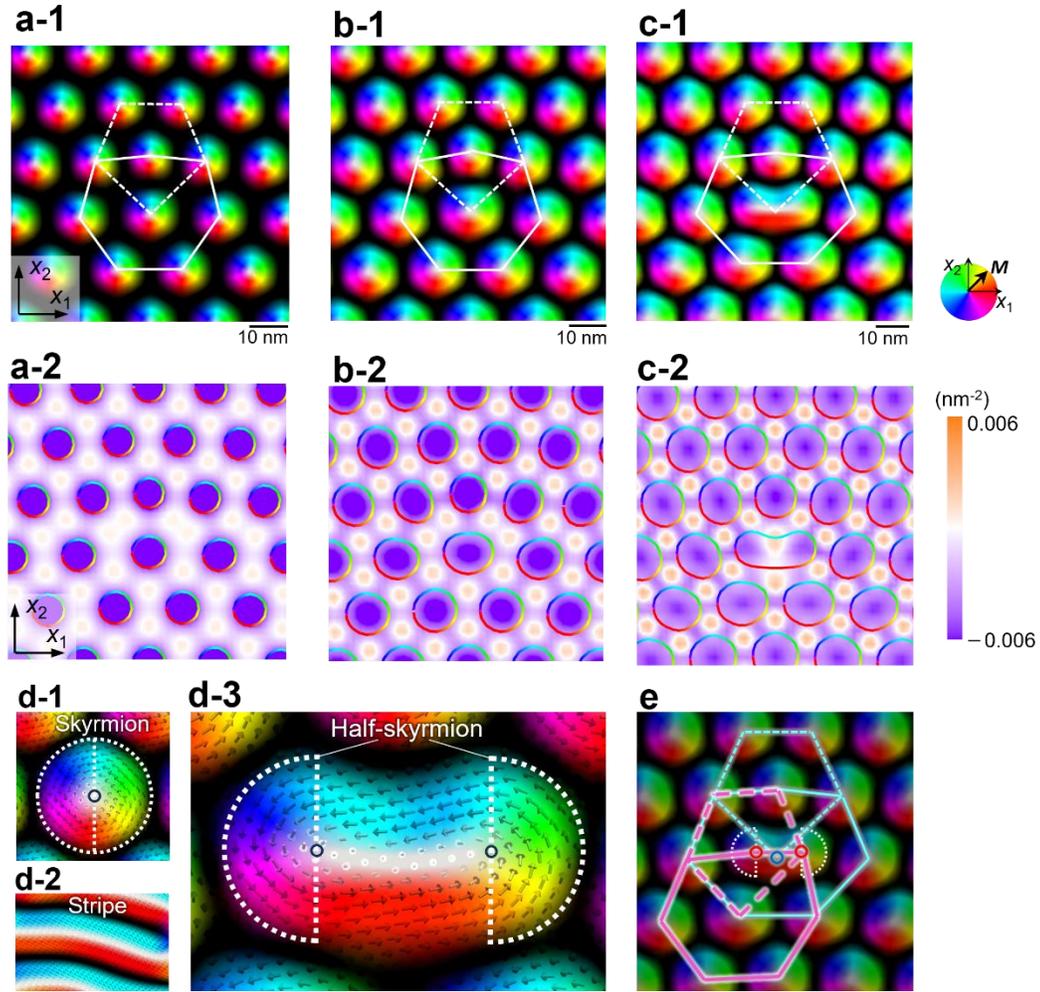

Figure 2. (a-1)-(c-1) Magnetic moment distribution near the dislocation core under different magnetic fields $H = 3.5\times10^5$, $2.0\times10^5$, and $1.0\times10^5$ A/m, respectively. (a-2)-(c-2) Pontryagin charge densities in the same area as (a-1)-(c-1). The colored lines represent the skyrmion shapes. (d-1)-(d-3) Magnetic moment configurations of a circular skyrmion, a stripe phase, and the elongated 7-fold skyrmion composed of two split half-skyrmions connected by a stripe phase region. The white dashed lines indicate the regions corresponding to half-skyrmions. (e) Lattice reconstruction associated with the topological transition. Blue lines denote the original 5-7 pair structure, whereas red lines indicate the reconstructed 5-7 pair considering the topological nature of half-skyrmions.

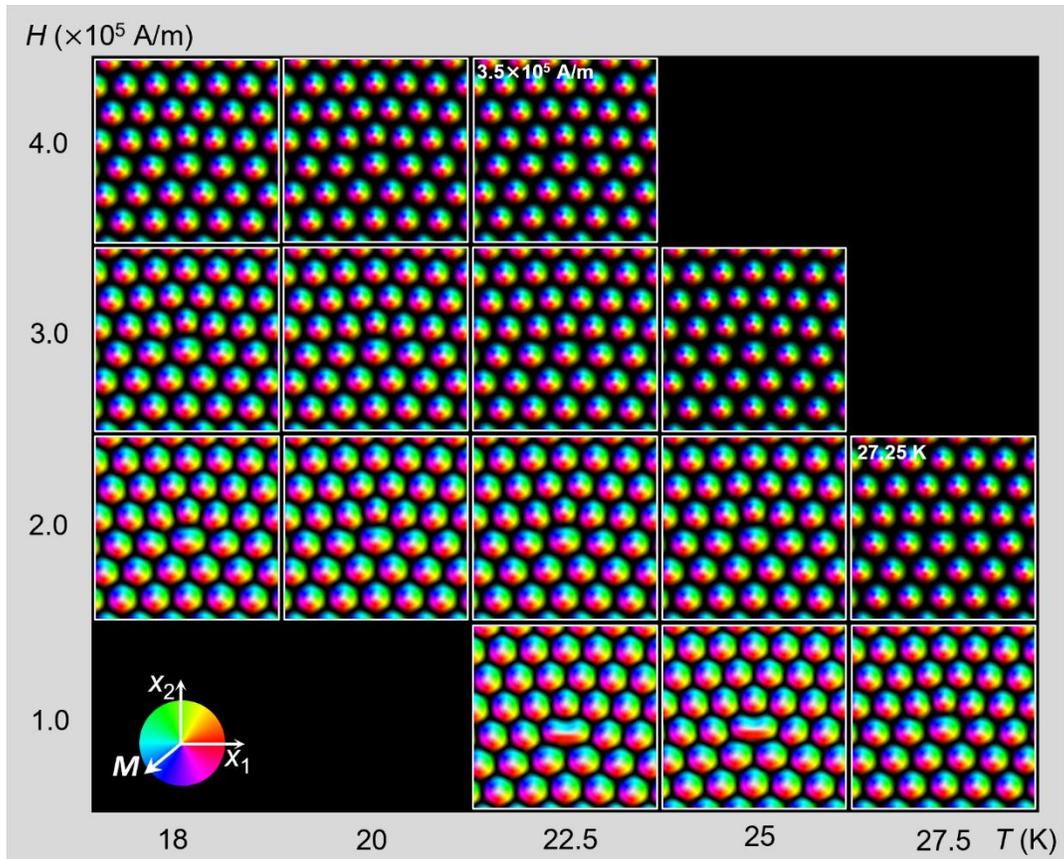

Figure 3. The magnetic fields-temperature *H-T* phase diagram of the skyrmion dislocation structure. The black region represents the conditions under which the skyrmion phase is unstable.

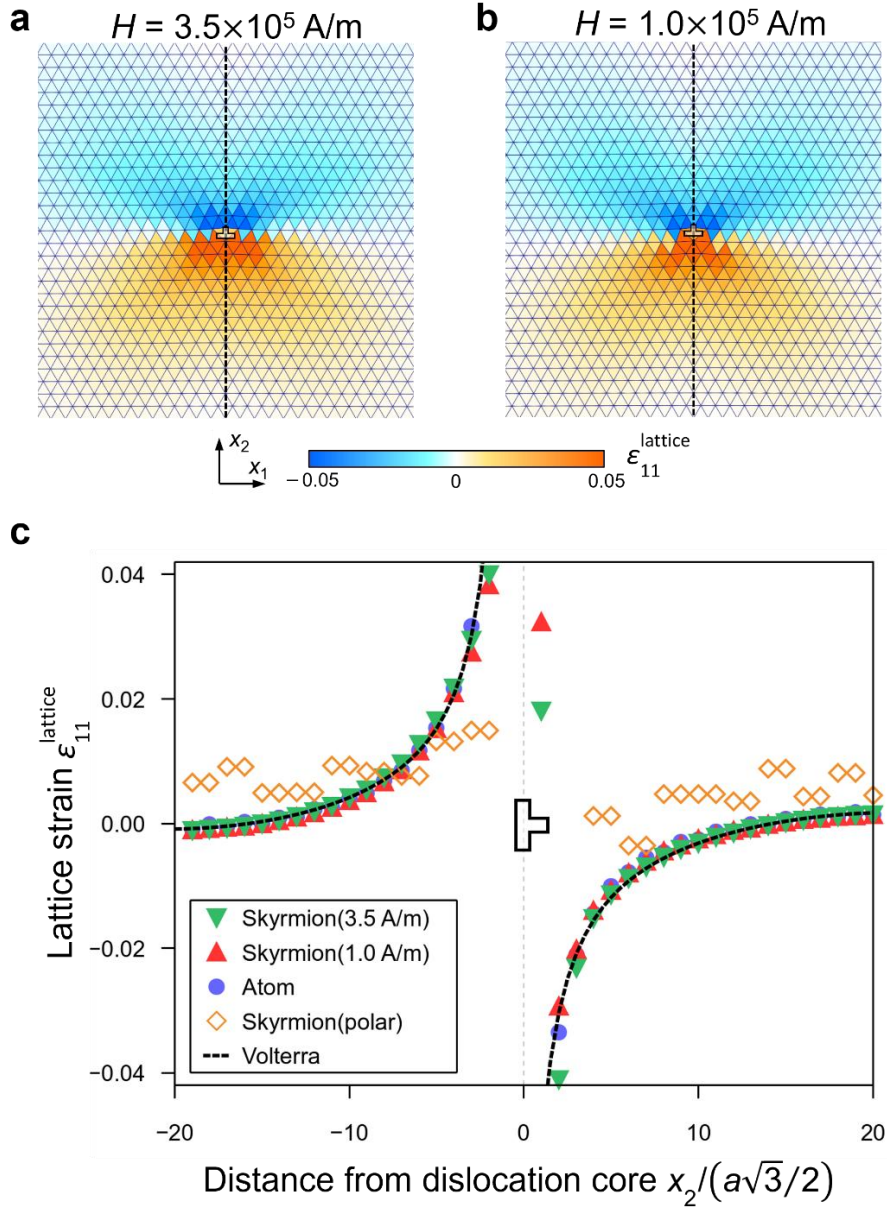

Figure 4. Lattice strain distribution around the skyrmion dislocation under (a) $H = 3.5 \times 10^5$ A/m, and (b) $H = 1.0 \times 10^5$ A/m. (c) Lattice strain as a function of the distance from the dislocation core on the centerline [dashed lines in (a)(b)] that intersect the dislocation cores. For comparison, values for an atomic crystal, polar skyrmions[45], and the theoretical solution of Volterra's law are shown. The distance is normalized by the lattice constant $a$ of each system.

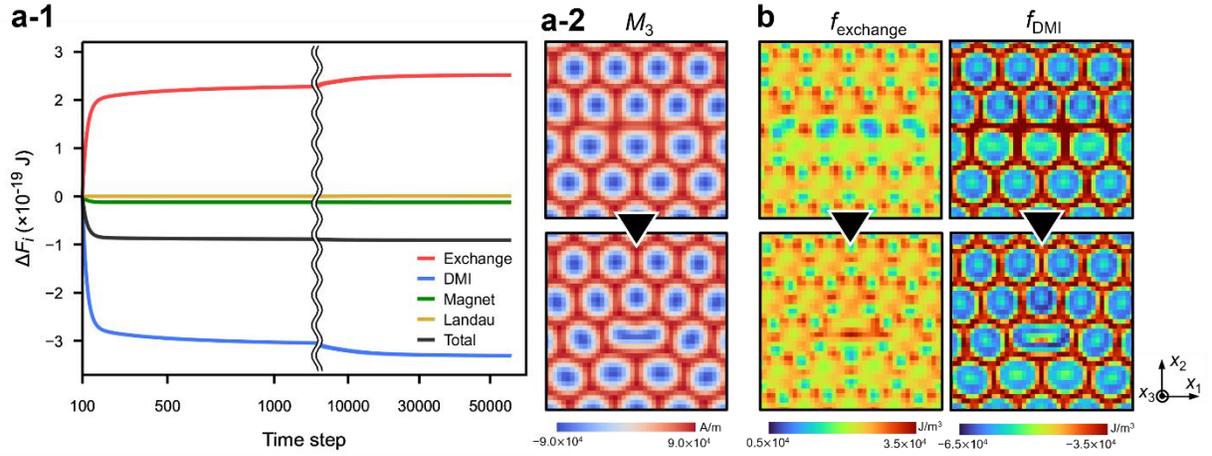

Figure 5. (a-1) Changes in each free energy component $\Delta F_i$ during the skyrmion deformation process, relative to the value at 100 steps. Note that DMI anisoropy energy and elastic energy are not shown because their variations are negligible. (a-2) Magnetic moment component $M_3$ distribution near the dislocation core before and after large skyrmion deformation (at 100 steps and at the final state, 55873 steps). (b) Distributions of DMI energy density and exchange energy density in the same area as (a-2).